\documentclass[fleqn,twoside]{article}
\usepackage{espcrc2}
\usepackage{epsfig}

\usepackage{graphicx}
\usepackage[figuresright]{rotating}

\newcommand{\AmS}{{\protect\the\textfont2
  A\kern-.1667em\lower.5ex\hbox{M}\kern-.125emS}}

\title{Implications of Confirmation of the LSND $\overline{\nu}_{\mu} \rightarrow 
        \overline{\nu}_e$ Oscillation Signal}

      \author{H. Ray\address[LANL]{Los Alamos National Laboratory \\
          Los Alamos, New Mexico 87545, U.S.A}%
        \thanks{current address : Fermi National Laboratory,
          MiniBooNE c/o M.S. 309,
          Batavia, IL 60510, U.S.A}
      }

\begin{document}

\begin{abstract}
Neutrino oscillations have been observed in solar and atmospheric neutrinos, and in the 
LSND accelerator experiment.  
The Standard Model cannot accommodate all three positive results.  The solar and atmospheric results have been
confirmed.  An oscillation signal seen by MiniBooNE will validate the oscillation 
signal seen by LSND.  The question then becomes one of refining the Standard Model to 
allow for these three results.  
Four theories which can accommodate all three oscillation observations are 
the existence of sterile neutrinos, CP(T) violation, the existence of variable
 mass neutrinos, and small Lorentz violations.  
The Spallation Neutron Source (SNS), located at Oak Ridge Laboratories, Oak Ridge, Tennessee, will 
provide an ideal site to test these hypotheses.  The SNS, due to turn on in 2008, will 
supply a high intensity neutrino source of known flux and energy spectrum.  This source permits 
experiments to probe the high $\triangle \mathrm{m}^2$ region for measurements, 
where a positive signal from MiniBooNE would lie.  
\vspace{1pc}
\end{abstract}

\maketitle

\section{Introduction}

Neutrino oscillations have been observed at three different mass scales.  Oscillations of 
solar neutrinos are seen as a deficit of $\nu_e$ from the sun.  These oscillations were 
first observed by the Homestake~\cite{Homestake}, SuperKamiokande~\cite{SuperK}, 
SAGE~\cite{SAGE}, and GALLEX~\cite{GALLEX} experiments, and later 
confirmed by SNO~\cite{SNO} and KamLAND~\cite{Kamland}.  
Solar oscillations occur at a $\triangle \mathrm{m}^2$ of 
approximately 8x$10^{-5} \mathrm{eV}^2$.  Oscillations in atmospheric neutrinos 
($\triangle \mathrm{m}^2$ $\approx$2x$10^{-3} \mathrm{eV}^2$) are seen as a deficit 
in the ratio of $\nu_{\mu}$ to $\nu_e$ (Kamiokande~\cite{Kamiokande}), 
and in a zenith angle discrepancy in the upward 
direction (SuperKamiokande).  This result has been confirmed by K2K~\cite{K2K}.  The 
third regime in which neutrino oscillations have been observed is at the mass scale of 
$\triangle \mathrm{m}^2$ $\approx 0.1 \rightarrow 10 \mathrm{eV}^2$.  This result was 
observed by the LSND~\cite{LSND} experiment, and is currently awaiting confirmation.

In the Standard Model there are three active neutrinos.  In this 
scenario a summation law holds such that $\triangle \mathrm{m_{12}}^2$ + 
$\triangle \mathrm{m_{23}}^2$  = $\triangle \mathrm{m_{13}}^2$.  
The three mass scales of observed oscillations do not follow this law.  Therefore, 
one of the results is faulty or there exists physics beyond the Standard Model.  

The solar and atmospheric oscillations have been observed and confirmed by several experiments.  
Experiments such as KARMEN~\cite{Karmen} and CHOOZ~\cite{CHOOZ}
have attempted to confirm the LSND result; so far these experiments 
have produced null results.  However, none of these experiments have fully explored all 
of the LSND allowed regions.  The MiniBooNE~\cite{MiniBooNE} 
experiment will make a definitive test of the LSND signal.

\subsection{MiniBooNE}
To confirm or refute the LSND result an experiment is needed which will cover all 
of the LSND allowed region in $\sin^2 2 \theta$-$\triangle \mathrm{m}^2$ space, 
with a similar L/E and different sources of systematic uncertainties.  This experiment is MiniBooNE.
The MiniBooNE experiment is covered elsewhere in full detail and will only briefly be described here 
~\cite{MiniBooNE}.

MiniBooNE is a fixed target experiment which directs an 8 GeV
proton beam onto a beryllium target.  Positive particles produced by this collision (primarily 
pions and kaons) are collimated by a magnetic focusing horn.  These particles enter a 
50 meter decay region where they decay in flight to produce a $\nu_{\mu}$ beam.  
The neutrinos then travel through approximately 490 meters of a dirt absorber 
before entering the MiniBooNE detector.  
The energy of these neutrinos is $\approx$ 700 MeV, providing an L/E of  
$\approx$ 0.8 m/MeV (compared to the LSND L/E of $\approx$1 m/MeV).  
The MiniBooNE detector is a spherical tank filled with 800 tons of pure mineral oil.  The 
inside of the tank is lined with PMTs, providing 10\% photocathode coverage.  
Different beam energy, beam duty cycle, and oil allow MiniBooNE drastically different 
systematic errors than those found at LSND.

\section{Physics of a Positive LSND Result}

The Standard Model cannot accommodate a positive oscillation result from all three sectors 
(solar, atmospheric, LSND).  Should MiniBooNE confirm the LSND result, the driving question 
in neutrino physics experiments will be determining the physics which would explain all three positive results.   
For brevity's sake only four theories will be touched on in this proceeding.

\subsection{Sterile Neutrinos}
The existence of additional neutrinos has been proposed.  These ``sterile'' neutrinos 
are weak isospin singlets and do not engage in weak interactions.  Sterile neutrino models 
are known as ``3+\emph{n}'', where 3 refers to the number of active neutrinos 
in the Standard Model and \emph{n} is the 
number of sterile neutrinos.  Models with one and two additional sterile neutrinos have been studied.  
Of the tested models, the 3+2 theory has the best fit to current oscillation data.  In these theories 
the high mass eigenstates are composed almost entirely of the sterile neutrino flavor eigenstate, 
while the lower mass states are composed of the active eigenstates.
~\cite{Sorrel03}~\cite{Cirelli04}

\subsection{Mass Varying Neutrinos}
All oscillation results may be accounted for if we permit the three active Standard Model neutrinos 
to have variable masses which depend on the value of a scalar field \emph{A}.  In this model 
sub-gravitational strength interactions between ordinary matter and \emph{A} naturally occur.  The value of 
\emph{A} (and thus the mass of the neutrino) will change depending on the presence or lack of matter.  
~\cite{Kaplan04}~\cite{Zurek04}

\subsection{Lorentz Violations}
A small Lorentz symmetry violation would explain all three positive oscillation results.  
Lorentz violations cause the oscillation to be dependent upon the direction 
of propagation, and thus would be easy to search for.  The size of the violation 
required to accommodate oscillation data lies in the range expected for effects 
at the Plank scale in the presence of an underlying unified theory of general relativity with the Standard Model.
 ~\cite{Kostel04}

\subsection{CPT Violation}
Finally, there could exist CP or CPT violation.  In the CPT model there is no need to introduce 
additional sterile neutrinos.  The oscillation results can be explained by allowing the 
$\triangle \mathrm{m}^2$ of $\nu$ (and thus the probability of oscillation) to differ from that of the  $\overline{\nu}$.  
This effect can be tested most cleanly by choosing an experiment which is capable of 
running in neutrino and anti-neutrino mode.

\section{Tests of New Physics Theories}
Several facilities may be used to test these new physics theories.  This proceeding will 
focus on tests which may be done using the MiniBooNE detector, and using a 
MiniBooNE-like detector at the SNS. 

\subsection{MiniBooNE}
The MiniBooNE detector is an excellent place to test these new physics theories.  
Mass varying neutrinos and Lorentz violations can be tested on the data used to 
perform the oscillation measurement.  For example, a possible indication of mass 
varying neutrinos would occur if MiniBooNE observes a 
positive oscillation result in the mass range excluded by BUGEY~\cite{BUGEY}
 ($\sim$0.1 to 0.25 $\mathrm{eV}^2$).  
The oscillation signal can be plotted as a function of sidereal position.  Any variations would 
indicate Lorentz violation.

Running MiniBooNE in anti-neutrino mode will provide a second data set which would be 
necessary to test for CP violations.  The construction of a second MiniBooNE tank located 
further upstream from the current detector would allow two measurements of the 
neutral current (NC) cross section.  A difference in the NC rate between the two detectors would indicate an 
oscillation into a sterile neutrino.

While it is possible to test these new physics theories at MiniBooNE, using a neutrino beam 
formed from the decay-in-flight of mesons allows for very tricky systematic errors due to 
beam flux and cross sections.  One way to avoid these systematics is to build a second detector at the 
MiniBoonNE site.  Another option is to 
move to a source which provides a decay-at-rest neutrino beam.  Such a source is currently 
under completion at the SNS.

\subsection{SNS}
The SNS ~\cite{SNS} is located at the Oakridge Laboratories, Oak Ridge, Tennessee, U.S.A.  
A one GeV proton beam, running with 700 ns pulses at a rate of 60 Hz, will impinge on a mercury target.  
The mercury will absorb the majority of the $\pi^-$ and $\mu^-$ before decay; the primary 
neutrino flux comes from $\pi^+$ and $\mu^+$ decay-at-rest.  The lifetime of the $\pi^+$ and $\mu^+$ 
relative to the beam window will provide good temporal separation of the $\nu_{\mu}$, $\nu_e$, and $\overline{\nu}_{\mu}$.  
Primary backgrounds to experiments at the SNS will come from cosmic rays and machine neutrons.

Currently under consideration are two MiniBooNE-like detectors : one would be $\sim$20 tons, located 20 meters 
from the neutrino source, and the other would be an 800 ton detector at a distance of 100 meters from the target
and in the backward direction relative to the proton beam.  These two detectors will be able to search for sterile neutrinos, 
test for CP/CPT and Lorentz violations, and test mass varying neutrino models.  In addition, the smaller detector 
may be filled with a liquid or aqueous nuclei such as carbon or hydrogen.  It could then be used to test cross sections which are 
vital for oscillation analyses at the SNS.

\section{Conclusions}
Should MiniBooNE confirm the LSND oscillation signal the next neutrino experiments 
will focus on determining which new physics model provides the mechanism 
for the oscillations, and on measuring the oscillation parameters precisely.  
Four possible models have been described in this proceeding; however, there 
are many permutations and many other theories proposed to explain the three 
positive oscillation results.  These new physics theories may be tested 
at several facilities, notably MiniBooNE and at the SNS.


\begin{thebibliography}{9}
\bibitem{Homestake} B. Cleveland \emph{et al},  Astrophys. J. {\bf 496}:505-526 (1998).
\bibitem{SuperK} F. Villante,  arXiv:hep-ph/9904563.
\bibitem{SAGE} J. Abdurashitov \emph{et al},  J. Exp. Theor. Phys. {\bf 95}:181-193 (2002).
\bibitem{GALLEX} W. Hampel \emph{et al},  Phys. Lett. B {\bf 447}:127-133 (1999).
\bibitem{SNO}  Q. Ahmad \emph{et al},  Phys. Rev. Lett. {\bf 89}:011302 (2002).
\bibitem{Kamland} K. Equchi \emph{et al},  Phys. Rev. Lett. {\bf 90}:021802 (2003).
\bibitem{Kamiokande} Y. Fukuda \emph{et al},  Phys. Lett. B {\bf 433}:9-18 (1998).
\bibitem{K2K} Y. Oyama,  arXiv:hep-ex/9803014.
\bibitem{LSND} A. Aguilar \emph{et al},  Phys. Rev. D {\bf 64} 112007 (2001).
\bibitem{Karmen} K. Eitel \emph{et al},  Nucl. Phys. Proc. Suppl. {\bf 778}:212-219 (1999).
\bibitem{CHOOZ} M. Apollonio \emph{et al},  Phys. Lett. B {\bf 466}:415-430 (1999).
\bibitem{MiniBooNE} See H. Tanaka, ``MiniBooNE'', these proceedings.

\bibitem{Sorrel03} M. Sorel, J. Conrad, and M. Shaevitz,  arXiv:hep-ph/0305255.
\bibitem{Cirelli04} M. Cirelli, G. Marandella, A. Strumia, and F. Vissani,  arXiv:hep-ph/0403158.
\bibitem{Kaplan04} D. Kaplan, A. Nelson, and N. Weiner,  arXiv:hep-ph/0401099.
\bibitem{Zurek04} K. Zurek,  arXiv:hep-ph/0405141.
\bibitem{Kostel04} V. Kostelecky and M. Mewes,  arXiv:hep-ph/0406255.
\bibitem{Mavro04} N. Mavromatos,  arXiv:hep-ph/0402005.

\bibitem{BUGEY} Y. Declais \emph{et al},  Nucl. Phys. B {\bf 434}:503-534 (1995).
\bibitem{SNS} F. Avignone \emph{et al},  J. Phys. G:Nucl. Part. Phys. {\bf 29} 2497-2668 (2003). \\
  http://www.phy.ornl.gov/workshops/nusns/
\end{thebibliography}
\end{document}